\documentclass[11pt]{article}
\usepackage{makeidx}  
\usepackage{amsmath}
\usepackage{amsfonts}
\usepackage{amssymb}
\usepackage{latexsym}
\usepackage{psfrag}
\usepackage{theorem}
\usepackage{fullpage}

\usepackage[usenames,dvipsnames]{color}

\newcommand{\NATURALS}{\ensuremath{\mathbb{N}}}

\newcommand{\E}{\ensuremath{\mathbb{E}}}

\renewcommand{\epsilon}{\varepsilon}

\renewcommand{\P}{\ensuremath{\mathbb{P}}}

\newtheorem{thm}{Theorem}
\newtheorem{cor}[thm]{Corollary}

\newtheorem{lemma}{Lemma}
\newtheorem{assumption}{Assumption}
\newtheorem{definition}{Definition}

{\theorembodyfont{\normalfont}
	
	\newtheorem{example}{Example}
	}

\newenvironment{bproof}{{\bf Proof.}}{\hfill\rule{2mm}{2mm}\\}
\newenvironment{lemmaproof}{{\bf Proof of Lemma.}}{\hfill\rule{2mm}{2mm}\\}
\newenvironment{proofsketch}{{\bf Proof Sketch.}}{\hfill\rule{2mm}{2mm}\\}

\title{Social Networks and Stable Matchings in the Job Market\footnote{A preliminary version will appear at the $5^{\text{th}}$ International Workshop on Internet and Network Economics, WINE 2009.}}

\author{
        Esteban Arcaute \\
        \ttfamily{arcaute@stanfordalumni.org}\\
        Stanford University\\
        Stanford, California, USA
            \and
        Sergei Vassilvitskii\\
        \ttfamily{sergei@yahoo-inc.com}\\
        Yahoo! Research\\
        New York, New York, USA
}
\date{\today}

\begin{document}

\maketitle

\begin{abstract}
	For most people, social contacts play an integral part in finding a new job. As observed by Granovetter's seminal study, the proportion of jobs obtained through social contacts is usually large compared to those obtained through postings or agencies. At the same time, job markets are a natural example of {\em two-sided matching markets}. An important solution concept in such markets is that of {\em stable matchings}, and the use of the celebrated Gale-Shapley algorithm to compute them. So far, the literature has evolved separately, either focusing on the implications of information flowing through a social network, or on developing a mathematical theory of job markets through the use of two-sided matching techniques.\\

In this paper we provide a model of the job market that brings both aspects of job markets together. To model the social scientists' observations, we assume that workers learn {\em only} about positions in firms through social contacts. Given that information structure, we study both static properties of what we call {\em locally stable matchings} (i.e., stable matchings subject to informational constraints given by a social network) and dynamic properties through a reinterpretation of Gale-Shapley's algorithm as myopic best response dynamics. \\

We prove that, in general, the set of locally stable matching strictly contains that of stable matchings and it is in fact NP-complete to determine if they are identical. We also show that the lattice structure of stable matchings is in general absent. Finally, we focus on myopic best response dynamics inspired by the Gale-Shapley algorithm. We study the efficiency loss due to the informational constraints, providing both lower and upper bounds. 
\end{abstract}

\section{Introduction}
When looking for a new job, the most often heard advice is to ``ask your friends." While in the modern world almost all of the companies have online job application forms, these are usually overloaded with submissions; and it is no secret that submitting a resume through someone on the inside greatly increases the chances of the application actually being looked at by a qualified person. This is the underlying premise behind the professional social networking site LinkedIn, which now boasts more than 40 million users \cite{wikipedia}. And, as pointed out by Jackson \cite{jacksonbook} has given a new meaning to the word `networking,' with Merriam Webster's Dictionary's
defining it as ``the cultivation of productive relationships for employment or business.''  

Sociologists have long studied this phenomenon, and have time and time again confirmed the role that social ties play in getting a new job. Granovetter's
seminal work \cite{granovetter,granovetterbook} headlines a long
history of research into the importance of social contacts in labor
markets. His results are striking, for example,
65 percent of managerial workers found their job through social
contacts. Interestingly, as early as 1951, a similar study about
textile workers \cite{myers} found similar results: more than 60
percent found their job through social contacts. Other studies (see,
e.g., \cite{rees,montgomery,datcher}) all echo
the importance of social contacts in securing a new position.

While there are numerous reasons that social ties play such an important role, one may think that the employers themselves would prefer to evaluate {\em all} candidates for a position before making a hiring decision.  This is in fact what happens in some segments of the job market.  For example, in most Western European countries, applicants
to positions within the government (public servants) pass through a
centralized selection process: government officials first gather all
relevant information, then proceed to match applicants to positions.
In the United States, the National Resident Match Program is a
significant example of a centralized selection matching mechanism.
Such centralized markets have been well studied in {\em two-sided
   matching theory}. Indeed, the NRMP is one of the most important
practical applications of the celebrated {\em stable matching problem}
in two-sided matching markets \cite{roth1984a}. For an overview of
two-sided matching markets, see \cite{rothsotomayor}.

However, the task of evaluating (and ranking) all possible candidates is often simply not feasible. Especially in today's economy, it is not rare to hear of hundreds of applicants for a position, obviously the vast majority cannot be interviewed, regardless of their qualifications. The recommendation by an employee thus carries extra weight in the decision process, precisely because it separates the specific application from the masses. (It can be argued that making good recommendations is also in the employee's best interest, but we will not be addressing incentive issues in this work.)

\paragraph{Model} - 
In this work, we propose a new model that bridges the rigorous analysis of the two-sided matching theory with the observations made by social network analysis. 
Specifically, we develop a model of job
markets where social contacts play a pivotal role; and then proceed to analyze it through the stable matching lens.

We integrate the usage of social contacts by allowing an applicant to apply {\em only} to jobs in firms employing her friends. 
Clearly this limitation depends on the underlying social graph. If it is a clique (that is everyone knows everyone else), the equilibrium behavior is exactly the same as in classical two sided matching. If it is disconnected, then some workers will never discover some firms, leading to potentially very inefficient matchings.  But even in well connected social graphs this limitation leads to behaviors not observed in the traditional model. For example, a firm may lose all of its workers to the competition, and subsequently go out of business (if no one is employed at a firm, no one can recommend others to join, thereby forcing it to close its doors). 

The model forces us to consider a setting
where job applicants have only partial information on job
opportunities. The main question we focus on in the paper is: how does the inclusion of such an informational constraint alter the model and predictions of traditional stable matching theory?

\paragraph{Our Contributions} - 
In traditional two-sided matching theory, a matching where no
worker-firm pair can find a profitable deviation is called {\em
  stable}. Analogously, we call our solution concept a {\em locally
   stable matching}, where the locality is qualified by the social network graph.  We study structural properties of locally stable
matchings by showing that, in general, the set of locally stable
matchings does not form a distributive lattice, as is the case for
global stable matchings. We  also show that, in general, it is
NP-complete to determine whether all locally stable matchings are also globally stable.  
Both of these results exploit a characterization of locally stable matchings in the special case of
matching one worker per firm, for particular rankings over workers and
firms.

We then turn our attention to dynamic analysis. We
consider how a particular interpretation of the classic Gale-Shapley
algorithm \cite{galeshapley} performs under such informational constraints; we
refer to our algorithm as the {\em local Gale-Shapley} algorithm.  We
first prove that, unlike the standard Gale-Shapley algorithm 
\cite{rothsotomayor}, the
existence of informational constraints implies
that the output of the algorithm is not independent of the order of
proposals.  Nevertheless, under weak stochastic conditions, we show
that the local Gale-Shapley algorithm converges almost surely, assuming the same particular rankings over workers and firms as before. 

Unlike the traditional Gale-Shapley algorithm, the algorithm in
the limited information case is highly dependent on the initial
conditions. To explore this further we define a minimal notion of efficiency,
namely the number of firms still in business in the outcome matching,
and quantify
the efficiency loss under various initial conditions. Specifically, we
show that
if an adversary chooses an initial matching, he can ensure that some firms
lose all of their workers; conversely there is a distribution on the
preference lists used by the firms
that guarantees that at least some firms remain in business,
regardless of the actions of the adversary.

We finish with a model that incorporates graph constraints into a two-sided matching market, where there is no restriction on the graph. As a preliminary result, we provide a full characterization of when locally stable matchings coincide with global stable matchings of the two-sided market. 

\subsection*{Related Work}

Our work touches on several threads of the literature. Most closely related is the work by Calv\'o-Armengol and Jackson \cite{calvo1,calvo2}. They consider how information dissemination through neighbors of workers on potential jobs can affect wage and employment dynamics in the job market. There are several key differences with our model. The most important one is that, in \cite{calvo1,calvo2}, there is no competition for job openings between workers. Unemployment is the result of a random sampling process and not of strategic interactions between workers and firms. Also, all workers learn {\em directly} about potential job openings with some probability, and {\em indirectly} through their social contacts, whereas in our model a worker can only learn about potential job openings through her social contacts.

Also related is the work by Lee and Schwarz \cite{interviewingschwarz}. The authors consider the stable matching problem in the job market where a costly information acquisition step (interviewing) is necessary for both workers and firms to learn their preferences. Once interviewing is over, the standard Gale-Shapley algorithm is used to calculate the matching of workers to firms. Although the authors use stable matching as their solution concept, and only partial information on jobs and candidates is available, their assumptions imply that the information available to workers and firms is {\em unchanged} throughout the matching phase. In that sense, their work is related to the equilibrium analysis performed in our model, but is dramatically different when considering the evolution of the job market during the actual matching phase.

There is a large body of literature on the impact of incomplete information in two-sided matching markets. Roth \cite{rothincomplete} studies that impact on strategic decisions. He concludes that phenomena arising in {\em all} equilibria of the complete information game need not happen in {\em any} equilibria of the corresponding incomplete information setting. Another example is the work of Chakraborty, Citanna and Ostrovsky \cite{interdependent}. They show how different assumptions on the information available to colleges when selecting aspiring students affect the existence of a stable matching mechanism.

In the study of uncoordinated two-sided matching markets, first introduced by Knuth in \cite{knuthuncoordinated}, agents on one side of the market {\em simultaneously} pick an agent on the other side of the market. Stable matchings correspond to Nash equilibria of the one-shot game. In \cite{uncoordinated}, the authors prove that the convergence time of best and better response dynamics can be exponential. Their (negative) results do not immediately apply to our model as, in the dynamics we consider, only {\em one} agent performs a best response at a time. 

In \cite{repeatedmatching}, the authors consider an experimental repeated matching market for workers and firms. They study how the underlying information structure on workers and firms' preferences affect the predictive power of the Gale-Shapley algorithm. Thus their interpretation of the Gale-Shapley algorithm as a predictor of {\em real} dynamics in information constrained repeated matching markets is related to our interpretation of the algorithm as the evolution of the job market under myopic best response dynamics.

Finally, in \cite{popular,mahdianpopular}, the authors consider the problem of matching applicants to job positions. A matching is said to be {\em popular} if the number of happy applicants is as large as possible. This notion is related to the notion of efficiency used in our paper, namely that of maximizing the number of firms in business.
 
\section{Definitions and Notation}
\label{sec:def}

Let $W$ be a set of workers, and $G = (W,E)$ be an undirected graph
representing the social network among workers. Let $F$ be the set of
firms, each with $k$ jobs, for some $k > 0$.  We are interested
in the case where there are as many workers as positions in all firms,
i.e., $|W|=n=k|F|$. Following standard notation, for a worker $w\in
W$, let $\Gamma(w)$ be the neighborhood of $w$ in $G$.

An assignment of workers to firms can be described by a function
mapping workers to jobs, or alternatively by a function mapping firms
to workers.
Following the definition from the two-sided matching literature, we
define both functions simultaneously.

We assume that some companies are better to work for than others, and
thus each worker $w$ has a strict ranking $\succ_w$ over firms such
that, for firms $f\neq f'$, $w$ prefers being employed in $f$ than in
$f'$ if and only if $f\succ_w f'$. Note however, that the ranking is
blind to the individual positions within a firm: all of the $k$ slots
of a given firm are equivalent from the point of view of a worker.

Similarly, each firm $f$ has a strict ranking $\succ_f$ over
workers.  We assume that all workers strictly
prefer being employed, and that all firms strictly prefer having all
their positions filled. For any worker $w$, in a slight abuse of
notation, we extend her ranking over
firms to account for her being unemployed by setting $f\succ_w w$ for
all firms $f$; in a similar way we extend the rankings of firms over
workers.

\begin{definition}[Matching]
\label{def:matching}
\begin{enumerate}
\item {\em Case 1: $k = 1$.} The function $\mu:\ W\cup F \rightarrow W\cup F$
         is a {\em matching} if the following conditions
         hold: (1) for all $w\in W$, $\mu(w)
         \in F\cup\{w\}$; (2) for all $f\in F$, $\mu(f) \in
         W\cup\{f\}$; and (3) $\mu(w) = f$ if and only if $\mu(f) = w$.
\item {\em Case 2: $k > 1$.} The function $\mu:\ W\cup F \rightarrow
   2^W\cup F$ is a {\em matching} if the following conditions hold: (1)
   for all $w\in W$, $\mu(w) \in F\cup\{\{w\}\}$; (2) for all $f\in F$,
   $\mu(f) \in 2^W\cup\{f\}$; (3) $\mu(w) = f$ if and only if $w\in
   \mu(f)$; and (4) $|\mu(f)| \leq k$.
\end{enumerate}

        We say that a matching $\mu$ is {\em complete} if:
        \[
                \bigcup_{f\in F} \mu(f) = W.
        \]

\end{definition}

Given a matching $\mu$ and a firm $f$, let $\min(\mu(f))$
be the {\em least preferred} worker employed by firm $f$ (w.r.t. firm $f$'s 
ranking) if
$|\mu(f)|=k$, and
$\min(\mu(f)) = f$ otherwise.

To study the notion of stable matchings, we adapt the usual
concept of a {\em blocking pair}.  Given the preferences of workers
and firms, a matching $\mu$, a firm $f$ and a worker $w$, we say that
$(w,f)$ is a {\em blocking pair} if and only if $f\succ_w \mu(w)$ and
$w\succ_f\min(\mu(f))$.  In other words, worker $w$ prefers firm $f$
to her currently matched firm; and firm $w$ prefers worker $w$ to its
least preferred current employee.

Instead of using the standard notion of stable matching from the
matching literature, we define a generalization that accounts for
the locality  of information. Recall that a (global) matching is said to be
{\em stable} if there are no blocking pairs. However, in our paper we assume that
{\em the workers can only discover possible firms by looking at their
friends' places of employment.}  This informally captures a
significant mechanism of information transfer: although there may exist a
firm $f$ that would make $(w,f)$ a blocking pair, if none of $w$'s
friends work at $f$, then it becomes much less likely that $w$ would
learn of $f$ on her own.  We have the following definition.

\begin{definition}[Locally Stable Matching]
\label{def:localmatching}
Let $G =
(W,E)$ be the social network over the set of workers $W$. We say that a matching
$\mu$ is a {\em locally stable matching} with respect to $G$ if, for
all $w\in W$ and $f\in F$,  $(w,f)$ is a blocking pair if and only if
$\Gamma(w)\cap\mu(f) = \emptyset$ (i.e., no workers in $w$'s social
neighborhood are employed by firm $f$).
\end{definition}

Observe that this definition severely limits the number of possible
firms that can form a blocking pair with $w$. In fact, if the maximum
degree in $G$ is $\Delta$, the number of such blocking pairs is at
most $\Delta$, which is constant in most social network models. This
may initially suggest that the number of locally stable matchings is
typically quite large. However, the set of firms that are
``visible'' to a worker $f$ changes over time as other blocking pairs in
the system are resolved! Also note that for a given worker $w$, the
set of other workers she is competing against depends on both the
social network $G$ (i.e., her neighbors), and the current
matching.

\begin{example}[Indirect Competition]
        Assume $k=2$ and $G$ is the path over $W = \{w_1,w_2,w_3,w_4\}$:  
$w_1-w_2-w_3-w_4$. Consider worker $w_4$. If $\mu(f_1) = \{w_3,w_4\}$ and 
$\mu(f_2) = \{w_1,w_2\}$, then $w_4$ can only see positions in $f_1$. However, 
since $w_2$ is adjacent to $w_3$, $w_2$ can see all position in $f_1$. Hence, if 
$w_2\succ_{f_1}w_3\succ_{f_1}w_4$, $w_2$ could get $w_4$'s position in $f_1$, 
leading to $w_4$ being replaced by $w_2$ even though $w_2\notin \Gamma(w_4)$.
\end{example}

In the remainder of the paper, we characterize static
properties of locally stable matchings, and then analyze
dynamics similar to the Gale-Shapley algorithm.

\section{Static Analysis}
\label{sec:static}

For $k=1$, when the preferences of workers and firms are strict, it is known that the set of global stable matchings is a {\em distributive lattice}. Several results on global stable matchings rely on the existence of the lattice as it allows us to navigate through the set of global stable matchings \cite{parallel}, and thus permits optimization problems over global stable matchings to be tackled.

In general, the distributive lattice structure of the set of global stable matchings is not present in the set of locally stable matchings. We first recall the Lattice Theorem (by Conway), and then show how, in general, it does not hold for locally stable matchings. The exposition of the Lattice Theorem is that found in \cite{rothsotomayor} (Theorem 2.16).

The two operations defining the distributive lattice over global stable matchings are as follows. Let $\mu$ and $\mu'$ be two matchings. Define the operation $\vee_W$ over $(\mu,\mu')$ as follows: $\mu\vee_W \mu':\ W\cup F \rightarrow W\cup F$ such that, for all $w\in W$, $\mu\vee_W \mu' (w) = \mu(w)$ if $\mu(w) \succ_w \mu'(w)$, and $\mu\vee_W \mu' (w) = \mu'(w)$ otherwise. For all $f\in F$, $\mu\vee_W \mu' (f) = \mu'(f)$ if $\mu(f) \succ_f \mu'(f)$, and $\mu\vee_W \mu' (f) = \mu(f)$ otherwise. We can similarly define $\wedge_W$ by exchanging the roles of workers and firms.

\begin{thm}[Lattice Theorem (Conway)]
\label{thm:conway}
	When all preferences are strict, if $\mu$ and $\mu'$ are stable matchings, then the functions $\lambda = \mu \vee_W \mu'$ and $\nu = \mu \wedge_W \mu'$ are both matchings. Furthermore, they are both stable.
\end{thm}

	In general, given strict preferences of workers and firms, Theorem~\ref{thm:conway} does not hold for the set of locally stable matchings. This is the content of the following example.

\begin{example}[Absence of Distributive Lattice]
\label{ex:lattice}
	We assume $k=1$, $W = \{w_1,w_2,w_3\}$ and $F = \{f_1,f_2,f_3\}$. Further, let the preferences of all workers be $f_1\succ f_2\succ f_3$. Similarly, let the preferences of all firms be $w_1\succ w_2\succ w_3$. Finally, assume the graph $G$ is the path with $w_2$ and $w_3$ at its endpoints.
	
	Let $\mu(w_i) = f_i$ (and $\mu(f_i)= w_i$). It is clear that $\mu$ is a 1-locally stable matching. Consider now $\mu'$ be such that $\mu'(w_1) = f_1$, $\mu'(w_2) = f_3$ and $\mu'(w_3) = f_2$ (and $\mu'(f_1) = w_1$, $\mu'(f_2)=w_3$ and $\mu'(f_3) = w_2$). The only blocking pair here is $(w_2,f_2)$, but $f_2 = \mu'(w_3)$ and $w_3\notin \Gamma(w_2)$. Hence $\mu'$ is a 1-locally stable matching.
	
	We now construct $\lambda = \mu\vee_W \mu'$. For all $i$, $\lambda(w_i) = f_i$. Now $\lambda(f_1) = w_1$ but $\lambda(f_2) = \lambda(f_3) = w_3$. Hence $\lambda$ is not a matching. 
\end{example}

While very useful in resolving optimization problems over global stable matchings, it is important to note that the absence of the distributive lattice has been previously observed when the preferences are not strict. Even the introduction of ties suffices for the distributive lattice, over {\em weak} global stable matching, to be absent \cite{roth1984a}. In \cite{POstructure}, the authors prove that the distributive lattice is present when ties are allowed for {\em strong} global stable matchings, but can be absent if the indifference takes the form of an arbitrary partial order. 

\paragraph{Assumption} - In the remainder of the paper we focus on a specific family of preferences over workers and firms. Under general preferences over workers and firms, the set of global stable matchings is not unique. When considering a two-sided matching market with global stable matching as its solution concept, uniqueness of global stable matching is a desirable property as it allows for sharp predictions of the outcome at equilibrium. In \cite{ClarkAR2006}, the author studies thoroughly the question and identifies a set of sufficient conditions on the preferences for the global stable matching to be unique. The set of preferences satisfying such conditions, called {\em aligned preferences}, have recently received attention in the economics literature \cite{MurielTR2009, SorensenAR2007}. 

In this paper we consider a subset of aligned preferences, where all workers share the same ranking over firms, and firms share the same ranking over workers. This assumption is made for technical reasons - we believe our results extend to the case of general aligned preferences.

\begin{assumption}
\label{as:preferences}
	There exist a labeling of the nodes in $W = \{w_1,\dots,w_n\}$ such that all firms rank workers as follows: $w_i\succ w_j$ if and only if $i<j$. Similarly, we assume there exists a labeling of the firms $F = \{f_1,\dots,f_{n_f}\}$ such that all workers rank the firms as follows: $f_i\succ f_j$ if and only if $i<j$.
\end{assumption}

We first show that, for $k=1$, the set of locally stable matchings is equivalent to the set of topological orderings over the partial order induced by $G$ and the labeling of the workers. 

\begin{thm}[Characterization of Locally Stable Matchings for $k=1$]
\label{thm:charac1}
	Let $G(W,E)$ be the social network over the set of workers. Let $D(W,E')$ be a directed graph over $W$ such that $(w_i,w_j)\in E'$ if and only if $i<j$ and $(w_i,w_j)\in E$.	Let $\mu$ be a complete matching of workers to firms. Construct the following ordering $\phi_{\mu}$ over $W$ induced by $\mu$: the $i^{\text{th}}$ node in the ordering is the node $w$ such that $\mu(w)=f_i$, i.e. $\phi_{\mu}(w)=i$. 
	
	The matching $\mu$ is a 1-locally stable matching if and only if $\phi_{\mu}$ is a topological ordering on $D$.
\end{thm}

\begin{bproof}
	Assume $\phi_{\mu}$ is a topological ordering on $D$. Then, for all $e=(w,w')\in E'$, $\phi_{\mu}(w)<\phi_{\mu}(w')$, which implies $\mu(w)\succ\mu(w')$. Recall that $(w,w')\in E'$ if and only if $w\succ w'$. Hence $\mu$ is locally stable.
	
	Assume $\phi_{\mu}$ is not a topological ordering over $D$. Then there exist $(w,w')\in E'$ such that $\phi_{\mu}(w)>\phi_{\mu}(w')$. This is equivalent to $\mu(w')\succ\mu(w)$. But recall that $w\succ w'$, thus $(w,\mu(w'))$ is a blocking pair, and $(w,w')\in E$, hence $\mu$ is not locally stable.
\end{bproof}

One important desirable property of the set of preferences in Assumption~\ref{as:preferences} is the uniqueness of the global stable matching - the highest $k$ workers are assigned to $f_1$, the next $k$ to $f_2$ and so on. There are several important corollaries to the characterization from Theorem~\ref{thm:charac1}. First, the set complete locally stable matchings can be exponentially large. Thus, by introducing informational constraints, the uniqueness property of global stable matchings under aligned preferences is, in general, lost under locally stable matchings.

\begin{cor}[Number of Locally Stable Matchings]
\label{cor:number}
	Assume $k=1$ and $G(W,E)$ is the star centered at worker $w_1$. Then there are $(n-1)!$ distinct locally stable matchings.
\end{cor}

\begin{bproof}
	Note that for all $i\neq 1$, $\Gamma(w_i) = \{w_1\}$. Also $w_1\succ w_i$. Hence, in any complete locally stable matching $\mu$, $\mu(w_1) = f_1$. Once the firm $w_1$ works for is set, the rest of the matching can be completed in any way as the only firms a worker $w_i\neq w_1$ can see are $f_1$ and her own firm. There are $(n-1)!$ such matchings, which proves the result.
\end{bproof}

The proof of Corollary~\ref{cor:number} assumes a specific topology for the social network $G$. It is thus interesting to ask whether there are specific properties of the social network $G$ that guarantee the existence of a labeling under which there is a unique complete locally stable matching. As shown in the next corollary, it is NP-complete to answer positively such question.

\begin{cor}
\label{cor:uniquenpc}
	Let $(k,G(W,E))$ be given. It is NP-complete to test if there is a labeling $\{w_1,w_2,\dots,w_n\}$ of the workers such that, if all firms rank the workers according to that labeling, the complete locally stable matching is unique.
\end{cor}

In order to prove Corollary~\ref{cor:uniquenpc}, we use Lemma~\ref{lem:topo}, a characterization of directed acyclic graphs with a unique topological ordering.  Lemma~\ref{lem:topo} can be found in Appendix~\ref{ap:technicallemmas}.

The proof of Corollary~\ref{cor:uniquenpc} is then as follows.

\begin{bproof}
	We prove the corollary by focusing on the case $k=1$. Since all firms use the same ranking over the set of workers, there is a unique global stable matching. Hence it suffices to show it is NP-complete to test whether there is a labeling of the workers such that there is a unique locally stable matching.
	
	By Theorem~\ref{thm:charac1}, there is a unique locally stable matching if and only if there is a unique topological ordering of $D(W,E')$, where $D$ is a directed acyclic graph such that $(w_i,w_j)\in E'$ if and only if $i<j$ and $(w_i,w_j)\in E$. By Lemma~\ref{lem:topo}, $D$ has a unique topological ordering if and only if its longest path is of length $n-1$. When that is the case, the path is $w_1w_2\dots w_{n-1}w_n$. But a path in $D$ is also a path in $G$, hence there is a unique locally stable matching if and only if the path $w_1w_2\dots w_n$ is in $G$. 
	
	Thus, for $k=1$, a given labeling of the workers is such that all locally stable matchings are global stable matchings when firms rank workers according to that labeling if and only if there is a labeling $\{w_1,\dots,w_n\}$ of the nodes such that $w_1w_2\dots w_n$ is a path in $G$. Hence, for $k=1$, our problem is equivalent to Hamiltonian path.
\end{bproof}

We note that, for general $k>1$, the proof of Corollary~\ref{cor:uniquenpc} can be adapted to show that a {\em sufficient} condition for the complete locally stable matchings to be unique is for $G$ to have a Hamiltonian path such that the top $k$ nodes are all visited first, then the next top $k$ nodes and so on. 

\begin{cor}
\label{cor:sufficientunique}
	Let $G(W,E)$ and $k>1$ be given. Assume nodes in $W$ are labeled such that, for all firms $w_1\succ \dots \succ w_n$. For all $0\leq i< n_f$, define $W_i = \{w_{ki+1},\dots,w_{k(i+1)}\}$. If $G$ has a Hamiltonian path such that, for all $0\leq i<j<n_f$, all nodes in $W_i$ are visited before all nodes in $W_j$, then the complete locally stable matching is unique.
\end{cor}

The proof is obvious if the path $w_1w_2\dots w_{n-1}w_n$ exists in $G$. To see why the result holds, it suffices to note that nodes in the same firm do not compete with each other as all positions in a given firm are equally valued by all workers.

Unfortunately, the condition from Corollary~\ref{cor:sufficientunique} is not necessary as we can see in the following example.

\begin{example}
\label{ex:notnecessary}
	Assume $k=2$ and $G$ being the path $w_4-w_2-w_1-w_3$. Note that the subgraph over $W_2$ is disconnected, hence the condition from Corollary~\ref{cor:sufficientunique} does not hold. Now note that $w_1$ is adjacent to both $w_2$ and $w_3$, hence $w_1$ is employed by $f_1$ and either $w_2$ or $w_3$ is employed by $f_2$. Since $w_1$ is employed by $f_1$ and $w_1$ is adjacent to $w_2$, $w_2$ competes directly with any employee of $f_1$ other than $w_1$. Hence $w_2$ is employed by $f_1$ which proves that the complete locally stable matching is unique.
\end{example}

\section{Algorithmic Questions}
\label{sec:algo}

We are now interested in finding locally stable matchings. Since all global stable matchings are locally stable matchings, we could simply use the Gale-Shapley algorithm \cite{galeshapley} to find a global stable matching. However, Gale-Shapley's algorithm requires the proposing side of the matching market to traverse its preference list in descending order. Here we assume that workers propose (i.e. that workers apply for jobs). Thus, in order to run Gale-Shapley's algorithm, all workers need to {\em know} about all possible positions, even those in firms not employing any of its neighbors. 

We are thus interested in decentralized algorithms that can find a locally stable matching. In this section we propose a decentralized version of Gale-Shapley's algorithm. Assumption~\ref{as:preferences} is again enforced in this section.  We first prove that our algorithm converges. Unlike the case without informational constraints, our algorithm does not always select the same locally stable matching. 

Recall that the Gale-Shapley algorithm is initialized by an empty matching \cite{rothsotomayor}. Since the empty matching is a locally stable matching, our algorithm requires to be initialized by a non-empty matching. We thus explore our algorithm's performance under adversarial initial complete matchings. We use the number of firms with no employees as a proxy for efficiency. We characterize the potential efficiency loss by providing upper and lower bounds on the number of firms with no employees.

\subsection{Local Gale-Shapley Algorithm}

One can interpret the Gale-Shapley algorithm from two-sided matching theory as a constrained version of {\em myopic best response dynamics} in the following way.

The dynamics proceed in rounds, which we index by $q\in\NATURALS$. Let $\mu^{(q)}$ be the matching at the beginning of round $q$. Let $w^{(q)}\in W$ be the {\em active} worker, where $w^{(q)}$ is sampled uniformly at random from $W$, and independently from previous rounds. We call such sampling process the {\em activation process}. Such activation process can be thought of as follows: assume all workers decide to explore employment opportunities according to a random clock with an exponential distribution with a given mean (the same mean for all workers). When the clock of $w_i$ ``sets off'', $w_i$ becomes active and looks for a better job. It is easy to see that the sequence of active nodes has the same distribution as taking independent uniform samples from $W$.

In myopic best response dynamics, $w^{(q)}$ would consider its current firm $\mu^{(q)}(w^{(q)})$ and compare it to the best firm $f$ it could be employed by given $\mu^{(q)}$ (i.e. the best firm where the worst employee was worse than $w$ given the matching $\mu^{(q)}$). If its current firm was better, it would pass. Else it would quit its job and get employed by $f$ (leading to a worker being fired, or an empty position being filled).

Gale-Shapley's algorithm is a constrained version of the above dynamics as it requires the active worker to consider the best firm it has not considered before (in other words it requires the active worker to remember what firms he has already failed to get a position at). 

\vspace{2mm}

We consider a local and decentralized version of the myopic best-response dynamics proposed above. We call it ``local Gale-Shapley'' algorithm. Instead of restricting the strategy space of the active worker using ``memory'' as in Gale-Shapley's algorithm, we restrict it using the graph $G(W,E)$ in the following way: $w^{(q)}$ compares its current firm in $\mu^{(q)}$ to the best firm that employs one of its neighbors in $G$ it could be employed by given $\mu^{(q)}$. An alternative way to describe the process is that the active node $w^{(q)}$ applies for a job at all the firms employing its neighbors that she strictly prefers to her current employer, and selects the best offer she gets (that offer might eventually be to stay at her current job).

More formally, the algorithm proceeds in rounds indexed by $q\in \NATURALS$. During round $q\geq0$:
\begin{itemize}
	\item the active worker $w^{(q)}$ is sampled, independently from previous rounds, uniformly at random from $W$.
	\item Next, $w^{(q)}$ applies to all firms she strictly prefers to her current employer,$\mu^{(q)}(w^{(q)})$.
	\item The active worker receives some offers:
	\begin{itemize}
		\item if at least one offer is received, $w^{(q)}$ quits her current employer and joins the best firm that sent an offer;
		\item if no offers are received, $w^{(q)}$ stays at her current job. 
	\end{itemize}
\end{itemize}

It is important to note that, unlike the Gale-Shapley algorithm, this variant of best-response dynamics can lead to a firm loosing all its employees. This is due to the assumption that positions within firms are ``visible'' to workers only through neighbors in $G$. Thus, if a firm has no employees, its job openings are not visible to any worker.

\begin{example}[Firm with no Employees]
\label{ex:fob}
Let $n=4$ and $k=2$. Thus there are four workers and two firms. Assume that $G=K_4$. Consider the following initial matching:
\[
 \mu^{(0)}(f_1) = \{w_3,w_4\}\ \text{and}\ \mu^{(0)}(f_2) = \{w_1,w_2\}
\]
in other words, the best company has the worst workers. Then if we activate workers $w_1$ and $w_2$ before activating $w_3$ or $w_4$, both $w_1$ and $w_2$ would quit $f_2$ and work for $f_1$, getting both $w_3$ and $w_4$ fired. In that setting, $f_2$ has no employees, and thus the process ends.
\end{example}

It is also important to understand the need of the activation process. Recall that the matching found by the Gale-Shapley algorithm is independent on the order of activation of the workers \cite{rothsotomayor}. When considering locally stable matchings, this is no longer the case even if the underlying graph is the complete graph. Let us reconsider Example~\ref{ex:fob}. 

\begin{example}
	Now consider the resulting matching when the activation sequence is as follows: $\{w_1,w_4,w_2,w_3\}$. First, $w_1$ leaves $f_2$ and gets a position at $f_1$. This makes $w_4 = \min(\mu^{(0)}(f_1))$ unemployed. Next, since we activate $w_4$, she gets the free position from $f_2$. Next $w_2$ leaves $f_2$ and gets a position at $f_1$, which results in $w_3$ loosing her job. Finally, $w_3$ gets the free position at $f_2$. Thus the resulting matching is now
	\[
		\mu(f_1) = \{w_1,w_2\},\ \text{and}\ \mu(f_2) = \{w_3,w_4\}
	\]
	which is a locally stable matching different from that obtained with the activation sequence in Example~\ref{ex:fob}.
\end{example}

\vspace{2mm}

An important question is whether this local decentralized version of best response dynamics converges as it is not immediately clear it can't cycle. This is the content of our first result.

\begin{thm}[Convergence of Local Gale-Shapley Algorithm]
\label{thm:convergence}
Let $G(W,E)$ and $\mu^{(0)}$ be given. Then the local Gale-Shapley algorithm started at $\mu^{(0)}$ converges almost surely to a locally stable matching.
\end{thm}

\begin{proofsketch}
First note that if the algorithm converges, it does so to a locally stable matching. This is a consequence to the action taken by the active node $w^{(q)}$: it will change firms if and only if $w^{(q)}$ participates in a blocking pair of $\mu^{(q)}$. Let us now give a proof sketch of convergence.

We can assume without loss of generality that $f_1$ has at least one worker in $\mu^{(0)}$. If it doesn't, then we can relabel the firms and call $f_1$ be best firm with at least one employee in $\mu^{(0)}$. The proof proceeds in stages. During the first stage, since all workers prefer $f_1$ to all other firms, $f_1$ will never loose any employees. Moreover, since $G$ is connected, after sufficiently many rounds, all positions in $f_1$ are filled. Note that all workers are ordered the same way in all firms. Thus the worst worker of $f_1$ is strictly improving (or, equivalently, its label is strictly decreasing), which proves, by a standard Lyapunov argument, that the set of workers employed by $f_1$ converges.

The second stage starts once the set of workers employed by $f_1$ converged. Let $f_i$ be the best firm with at least one worker after the set of workers in $f_1$ converged. Then all workers not employed by $f_1$ strictly prefer $f_i$ to all other firms that have at least one worker. Hence, by an argument similar to that establishing convergence of the set of workers employed by $f_1$, the set of workers employed by $f_i$ converges. 

We can then repeat the same argument subject to the set of workers in $f_1$ and $f_i$ having converged, which proves the result.
\end{proofsketch}

\subsection{Worst Case Efficiency}

An important result in stable matching theory is the characterization of the matching obtained when using the Gale-Shapley algorithm. As we showed in Example~\ref{ex:lattice}, in general, the distributive lattice over the set of locally stable matchings is absent. Hence a result similar to that of the traditional stable matching literature seems unlikely. 

In this subsection we consider the following question. Given that firms can go out of business when running the local Gale-Shapley algorithm, can we measure the quality of matchings selected by the algorithm. Since we assume that workers are only aware of positions within firms employing her neighbors, we explore the previous question assuming a given initial complete matching $\mu^{(0)}$.

We consider the following setting. An adversary observes $G(W,E)$ (but not the ranking over workers used by firms) and produces a probability distribution ${\cal P}_M$ over initial matchings. The ranking of workers (possibly taken from a distribution) is then revealed, a sample from ${\cal P}_M$ is taken to produce $\mu^{(0)}$; and the local Gale-Shapley algorithm run. 

To compare the efficiency of different final matchings we simply look at the total number of firms losing all of their employees and subsequently going out of business. One can easily imagine more intricate notions of efficiency, our point here is that even in this austere model, the power of the adversary is non-trivial. 

\subsubsection{The power of the adversary}
We first show that even without knowing the relative rankings of the individual workers, the adversary is powerful enough to force some firms to go out of business.

\begin{thm}[Lower Bound on Firms]
\label{thm:lbfirms}
	Let $G(W,E)$ be given. Let $\Delta$ be its maximum degree, and $M$ a maximum matching in $G$. Then there exist a probability distribution ${\cal P}_M$ over complete assignment matchings such that
	\[
		\E[N_{\text{fob}}] \geq \left\lfloor \frac{|M|}{k(2\Delta)} \right\rfloor \frac{1}{2^k k!(2\Delta-1)^k}
	\]
	where $N_{\text{fob}}$ is the number of firms going out of business; and the expectation is taken both over the distribution ${\cal P}_M$ and over the activation process.
	
	Further, one can find ${\cal P}_M$ in time polynomial in $n$.
\end{thm}

	We provide a constructive proof of Theorem~\ref{thm:lbfirms}. First, we define a special type of matching in a graph $G$.
	
\begin{definition}[Independent Matching]
\label{def:independentmatching}
	Let $G(W,E)$ be an undirected graph. We say that $M\subseteq E$ is as {\em independent matching} if
	\begin{enumerate}
		\item $M$ is a matching; and
		\item the vertices matched by $M$ are an independent set of the subgraph $G'(W,E\setminus M)$.
	\end{enumerate}
\end{definition}

Equivalently, a matching is independent if, for any two nodes $u$ and $v$ matched in $M$ (but not necessarily matched to each other), $u\in \Gamma(v)$ if and only if $(u,v)\in M$.

In order to prove Theorem~\ref{thm:lbfirms}, we next prove the following lemma.

\begin{lemma}
\label{lem:independentmatching}
	Let $M'$ be an independent matching of $G$. Then there is an algorithm that constructs a distribution ${\cal P}_M$ in time $O(n)$ such that
	\[
		\E[N_{\text{fob}}] \geq \left\lfloor \frac{|M'|}{k} \right\rfloor\frac{1}{2^k k!(2\Delta-1)^k}
	\]
	where the expectation is taken both over the distribution ${\cal P}_M$ and over the activation process.
\end{lemma}

\begin{lemmaproof}
	Assume $M'$ has at least $k$ edges, and let $M' = \{e_1,\dots,e_{m'}\}$.
	
	For $\ell = 0$ to $\left\lfloor \frac{|M'|}{k} \right\rfloor-1$, consider the edges $\{e_{k\ell+1},\dots,e_{k(\ell+1)}\}$. For each edge $e_{k\ell + j}$, call $u_{k\ell + j}$ and $v_{k\ell + j}$ its endpoints. Since $M'$ is a matching, all such vertices are distinct. Consider a binary string of length $k$. Sample one such string $S = s_1\dots s_k$ uniformly at random. 
	
	For each $1\leq j \leq k$, set $\mu(u_{k\ell + j}) = f_{2\ell +1}$ and $\mu(v_{k\ell + j}) = f_{2(\ell+1)}$ if $s_j=0$. Otherwise, set $\mu(u_{k\ell +j})=f_{2(\ell+1)}$ and $\mu(v_{k\ell +j})=f_{2\ell+1}$.
	
	For all the remaining workers, assign all of the remaining positions in the firms in any order.
	
	Note that using the algorithm described above, all of the positions in the top $2\left\lfloor \frac{|M'|}{k} \right\rfloor$ firms are assigned to workers matched in $M'$. The following is a key fact derived from $M'$ being an independent matching:
	
	\noindent {\bf Key fact}: Let $(u,v)\in M'$ be one of the edges within the first $\left\lfloor \frac{|M'|}{k} \right\rfloor$ edges of $M'$. Let $f_{\max}$ be the best firm employing one of $u$ or $v$'s neighbors. Then $\mu(u)\succ f_{\max}$ and $\mu(v)\succ f_{\max}$. 
	
	The key fact follows from the nodes in $M'$ being only adjacent to nodes not in $M'$.
	
	Without loss of generality, assume that for all $\ell$ and $j$, $u_{k\ell +j}\succ v_{k\ell +j}$. Consider now firms $f_{2\ell +1}$ and $f_{2(\ell +1)}$. Note that
	\[
		\P[\forall 1\leq j\leq k,\ \mu(v_{k\ell +j})\succ \mu(u_{k\ell +j})] = \frac{1}{2^k}
	\]
	Assume further without loss of generality that $v_{k\ell +j}\succ v_{k\ell +i}$ for $j>i$. Then, if we activate $u_{k\ell +1}$ before all of the other nodes in 
	\[
		U_{k\ell +1} = \Gamma(u_{k\ell + i})\cup \Gamma(v_{k\ell + i})\cup\bigcup_{i=2}^k[u_{k\ell + i}\cup \Gamma(u_{k\ell + i})\cup \Gamma(v_{k\ell + i})]
	\]
	since $v_{k\ell +1}$ is $f_{2\ell +1}$ worst worker, $v_{k\ell +1}$ would loose its job to $u_{k\ell +1}$. This happens with probability 
	\begin{align*}
		\frac{1}{|U_{k\ell +1}|+1} &\geq \frac{1}{1+(2\Delta-2)+\sum_{i=2}^k|u_{k\ell + i}\cup \Gamma(u_{k\ell + i})\cup \Gamma(v_{k\ell + i})|}\\
															 &\geq \frac{1}{2\Delta-1+\sum_{i=2}^k (2\Delta-1)}\\
															 &= \frac{1}{k(2\Delta-1)}
	\end{align*}
	
	Conditioned on this event, the probability that $v_{k\ell +2}$ would loose its job to $u_{k\ell +2}$ is equal to 
	\[
		\frac{1}{|U_{k\ell +2}|+1} \geq \frac{1}{(k-1)(2\Delta-1)}
	\]
	
	More generally, the probability that the worst $i$ workers in firm $f_{2\ell +1}$ loose their job to the workers they were matched in $M'$ is at least
	\[
		\frac{1}{2^k}\prod_{j=1}^i\frac{1}{(k-j+1)(2\Delta-1)}
	\]
	and thus the probability that $f_{2(\ell+1)}$ goes out of business (because of all its workers going to work for firm $f_{2\ell+1}$) is at least
	\[
		\frac{1}{2^k k!(2\Delta-1)^k}.
	\]
	
	The result follows by linearity of expectation.	
\end{lemmaproof}

The proof of Theorem~\ref{thm:lbfirms} proceeds then as follows.

\begin{bproof}
	Note that, given an independent matching, the algorithm from Lemma~\ref{lem:independentmatching} takes $O(n)$ time. However, as we will show in Lemma ~\ref{lem:nphmatching}, finding the maximum independent matching is NP-hard. Here we give a simple algorithm that achieves a $2\Delta$ approximation. Let $M$ be a maximum matching in $G$. $M$ can be calculated in polynomial time. To get an independent matching, start from an edge $e$ in $M$ and remove all edges in $M$ that are adjacent to a neighbor of either endpoint of $e$. Doing so, we remove at most $2\Delta$ edges from $M$. By repeating this greedy algorithm (that takes $O(|E|)$ time) we create an independent matching $M'$ of size at least $|M|/(2\Delta)$.
	
	The previous fact, together with Lemma~\ref{lem:independentmatching}, proves the result.
\end{bproof}

Finally, note that the lower bound in Theorem~\ref{thm:lbfirms} does not use a maximum independent matching, but instead approximates one in order to achieve efficiency. This is due to the NP-hardness of finding a maximum independent matching. See Lemma~\ref{lem:nphmatching} in Appendix~\ref{ap:technicallemmas}.

An important observation is that not only does the adversary force some firms to go out of business, but he controls the identities of these firms. Thus, if we measure efficiency by the identity of the firms of the positions filled in a matching, Theorem~\ref{thm:lbfirms} provides a lower bound on the efficiency loss of the local Gale-Shapley algorithm (under adversarial initial conditions).

\subsubsection{The power of the social planner}

Given the lower bound from Theorem~\ref{thm:lbfirms} on the expected number of firms going out of business, we can ask the following question: can similar guarantees be proven if a social planner had full control over the ranking used by firms? More precisely, given $G(W,E)$, if the ranking over workers used by firms was a sample from a random variable, can the social planner guarantee, in expectation, a minimal number of firms that will not go out of business {\em regardless of the power given to the adversary}? The following theorem answers positively that question.

\begin{thm}[Upper Bound on Firms]
\label{thm:ubfirms}
	Let $G(W,E)$ be given. There exist a probability distribution over the ranking used by firms such that
	\[
		\E[N_{\text{fob}}] \leq n_f-\left\lceil \frac{|I|}{k} \right\rceil
	\]
	where $I$ is a maximum independent set of $G$ ($n_f$ is the number of firms and $k$ the number of positions at each firm)
\end{thm}

\begin{bproof}
	Let $I$ be a maximum independent set of $G$. Consider the following distribution over rankings used by firms. With probability one, assign the top $|I|$ workers to the nodes in $I$. Assign the rest arbitrarily. 
	
	Let $w\in I$ be given. Since $I$ is an independent set, all workers that are better ranked than $w$ are not neighbors of $w$. Thus, regardless of the initial assignment matching picked by the adversary, $w$ can lose its job to a worker $w'\succ w$ only if one of $w'$'s neighbors, say $y$ is employed by the same firm as $w$. Note that $w'\succ w$ implies $w'\in I$, and hence $y$ is not in $I$. It follows that $w\succ y$, and hence the firm employing $w$ would first fire $y$ to hire $w'$. We just proved that $w$ cannot lose its job $w'$. 
\end{bproof}

Note that, just as in Theorem~\ref{thm:lbfirms} we were able to identify the firms forced out of business (the top firms) but not the unemployed workers, in Theorem~\ref{thm:ubfirms} we are able to identify the workers that are going to be employed (the top employees), but not which firms will remain in business.

\subsubsection{Discussion}
We have now shown that neither the adversary, nor the social planner have all the power --- we can reinterpret the results above as a game between these two players. The game proceeds as follows, the adversary picks the initial assignment matching (possibly random), and the social planner chooses the ordering on the workers (possibly random). Once they both pick an action we run the local Gale-Shapley algorithm.

Theorem~\ref{thm:lbfirms} then states that, even if the social planner knows the probability distribution selected by the firm adversary, there is a probability distribution over initial assignments that the firm adversary can use such that, in expectation, at least some number of firms go out of business.

Theorem~\ref{thm:ubfirms} states the converse: Even if the firm adversary knows the probability distribution selected by the social planner, there is a deterministic ordering of the workers such that at least some number of workers will never loose their job.

We note that by looking at the number of firms going out of business, we have used a very minimal notion of efficiency. It is not hard to imagine more complex notions which may take into account the relative rankings of the firms going out of business or workers remaining unemployed. We further note that  for dense graphs, where the size of the independent set, and the independent matchings are quite small, our bounds are quite loose. Our main contribution here is not the precise bound on $N_{fob}$, although that remains an interesting open question, but rather the fact that the adversary has non-trivial power, and the initial matching plays a pivotal role in determining the final outcome. 

\section{Conclusions}
\label{sec:conclusion}
In this work we have introduced a new model for incorporating social network ties into classical stable matching theory. Specifically, we show that restricting the firms willing to consider a worker only to those employing his friends has a profound impact on the system. We defined the notion of locally stable matchings and showed that while a simple variation of the Gale-Shapley mechanism converges to a stable solution, this solution may be far from efficient; and, unlike in traditional Gale-Shapley, the initial matching plays a large role in the final outcome. In fact, if the adversary controls the initial matching, he can force some firms to be left with {\em no} workers in the final solution. 

The model we propose is ripe for extensions and further analysis. To give an example, we have assumed that as employees leave the firm, it may find itself with empty slots that it cannot fill (and go out of business). However, this is precisely the time when it can start looking actively for workers, by advertising online, recruiting through headhunters, etc. This has the effect of it becoming visible to the unemployed workers in the system. Understanding the dynamics and inefficiencies of final matchings under this scenario is one interesting open question. 

\subsection*{Acknowledgments}
We would first like to thank Ramesh Johari for many discussions on the model studied and its static analysis. We also want to thank David Liben-Nowell for the discussions that lead to the proofs of Theorems 7 and 8. Finally, we would like to thank Ravi Kumar and Matt Jackson for many useful discussions.\\
This research was supported by the National Science Foundation and by the Defense Advanced Research 
Projects Agency under the ITMANET program.

\vspace{5mm}

\bibliographystyle{abbrv}
\bibliography{matchings}

\section*{Appendix} 

\appendix

\section{General Model}
\label{ap:general}

In this appendix we consider a general framework to include graph constraints in two-sided matching markets with stable matching as solution concept. For ease of exposition we focus on the stable marriage problem, and assume agents have ordinal preferences. In this appendix Assumption~\ref{as:preferences} is not assumed to hold true. \\
More formally, we assume $n$ men $M$ and $n$ women $W$ participate in a social network $G(M\cup W,E)$, where $uv\in E$ represents a social connection between $u$ and $v$. For each man $m$ (and woman $w$), there is a ranking $\succ_m$ (respectively $\succ_w$) over all women $W$ (resp. men $M$) such that $w\succ_m w'$ indicates that $m$ prefers being matched to $m$ than to $m'$. 

The definition of matching is carried over from case 1 in Definition~\ref{def:matching} where we replace $F$ by $M$. The definition of locally stable matching is a straight forward generalization identical to that in Definition~\ref{def:localmatching}. For a given subset $S\subseteq M\cup W$ and a matching $\mu$, we call $\mu(S)\subseteq M\cup W$ the subset of men and women matched in $\mu$ to a men or women in $S$ (i.e. $x\in \mu(S)$ if and only if $\mu(x) \in S$ and $x\neq \mu(x)$).

\begin{definition}[Locally Stable Matching]
\label{def:localmatchinggeneral}
Let $G =(M\cup W,E)$ be the social network over the set of men $M$ and women $W$. We say that a matching
$\mu$ is a {\em locally stable matching} with respect to $G$ if, for
all $w\in W$ and $m\in M$,  $(w,m)$ is a blocking pair if and only if
$m\notin\Gamma(w)\cup\mu(\Gamma(w))$ and $w\notin \Gamma(m)\cup\mu(\Gamma(m))$ (i.e., $m$ (resp. $w$) is not directly connected to $w$ (resp. $m$) or matched to a neighbor of $w$ (resp. neighbor of $m$)). 
\end{definition}

We can now see that the case of the job market (with one position per firm) corresponds to the stable marriage problem where the social network is restricted to have edges between nodes in one side of the market only. Notice that in Section~\ref{sec:static}, we focused on complete matchings $\mu$. The reason was simple: the lack of edges between the two sides of the market made the empty matching a locally stable matching. Note that in this general framework, the empty matching is not de-facto locally stable: if $(m,w)\in E$ and $m$ (resp. $w$) prefers being matched to $w$ (resp. $m$) than being unmatched, then the empty matching is not locally stable. 

We now provide the main result of this appendix. We give necessary and sufficient conditions on $G$ under which the set of locally stable matchings coincides with that of global stable matchings {\em for all preferences}. This result is relevant for online social networks: the social network is observable, and membership of a node to either side of the market is available; preferences are hard if at all possible to infer. 

\begin{thm}
\label{thm:gsmlsm}
	Let $G(M\cup W,E)$ be given. Given preferences $\succ = (\succ_x;\ x\in M\cup W)$, we denote by ${\cal M}_{\succ}$ the set of global stable matchings with respect to $\succ$. Similarly, we denote by ${\cal M}_{\succ}(G)$ the set of locally stable matchings with respect to $\succ$ and $G$.
	
	For all $\succ$, ${\cal M}_{\succ}(G) = {\cal M}_{\succ}$ if and only if $K_{M,W}\subseteq G$, where $K_{M,W}$ is the complete bipartite graph with $M$ and $W$ as the partitions.
\end{thm}

\begin{bproof}
	From Definition~\ref{def:localmatchinggeneral} it is easy to see that $K_{M,W}\subseteq G$ implies that all locally stable matchings are global stable matchings as, for all $m$ (resp. $w$), $W \subseteq\Gamma(m)$ (resp. $M \subseteq\Gamma(w)$). We prove the direct direction by counter positive.
	
	Assume $K_{M,W}$ is not a subgraph of $G$, i.e. there exists $m_0\in M$ and $w_0\in W$ such that $(m_0,w_0)\notin G$. We now construct a set of preferences $\succ$ such that there exist $\mu\in{\cal M}_{\succ}(G)$ and $\mu\notin{\cal M}_{\succ}$. Consider a labeling of the set of men and women such that $M = \{m_i;\ 0\leq i<n\}$ and $W = \{w_i;\ 0\leq i<n\}$. We assume all men (resp. all women) share the same preferences $\succ_M$ (resp. $\succ_W$): $w_i \succ_M w_j$ if and only if $i>j$ (resp. $m_i \succ_W m_j$ if and only if $i>j$). 
	
	Under such preferences, the only global stable matching $\mu$ is such that, for all $0\leq i<n$, $\mu(w_i) = m_i$. Consider now the matching $\mu'$: for all $0<i<n$, $\mu'(w_i)= m_i$ and $\mu'(m_0) = m_0$ and $\mu'(w_0) = w_0$. In other words, $\mu$ and $\mu'$ coincide over $M\cup W \setminus \{m_0,w_0\}$ and leave $m_0$ and $w_0$ unmatched. 
	
	To see that $\mu'$ is in ${\cal M}_{\succ}(G)$, it suffices to prove that neither $m_0$, nor $w_0$ are in a blocking pair. First, note that all men (resp. women) other than $m_0$ (resp. $w_0$) prefer their match in $\mu'$ to $w_0$ (resp. $m_0$). Thus the only possible blocking pair $m_0$ could participate in is $(m_0,w_0)$. But $w_0\notin \Gamma(m_0)\cup\mu'(\Gamma(m_0))$, hence $(m_0,w_0)$ is not a blocking pair for $\mu'$ in $G$, which proves the result.
\end{bproof}

\section{Technical Lemmas}
\label{ap:technicallemmas}

In this appendix we state and prove two technical lemmas used in the paper. Lemma~\ref{lem:topo} is a necessary and sufficient condition for a directed acyclic graph to have a unique topological ordering. Lemma~\ref{lem:nphmatching} is a complexity result regarding finding maximum independent matchings.

\begin{lemma}[Topological Orderings]
\label{lem:topo}
	Let $D(V,E)$ be a directed acyclic graph over $n>1$ nodes. $D$ has a unique topological ordering if and only if the longest path in $D$ has length $n-1$.
\end{lemma}

\begin{lemmaproof}
	By induction on $n$. Base case: $n=2$. If $D$'s longest path is of length strictly less than $n-1 = 1$, then $D$ is the empty graph and any ordering of the nodes is a topological ordering. Since $n=2$, there are $2!=2$ such orderings. If $D$'s longest path is of length $n-1=1$, then $D$ has an edge between its two nodes, and the topological ordering is unique.
	
	Assume the inductive step holds for $n>1$. Assume $|V|=n+1$. We consider two cases.\\
	Case 1: there is a unique node $v$ in $D$ with indegree of zero. Then there is a path from $v$ to any other node in $V$. Hence $v$ is the first node in any topological ordering of $D$. Further, in any such $D$, all longest paths start from $v$. Hence the length of the longest path in $D$ is equal to 1 plus the length of the longest path in $D_v$, where $D_v$ is the subgraph of $D$ on $V\setminus\{v\}$. Thus $D$'s longest path is of length $n+1-1$ if and only if $D_v$'s longest path is of length $n-1$. Also, any topological ordering of $D$ consists of $v$ being the first node, and any topological ordering on $D_v$. Hence $D$ has a unique topological ordering if and only if $D_v$ has a unique topological ordering. Note that $D'$ has $n+1-1=n$ nodes, hence we apply the induction hypothesis.
	
	Case 2: there are at least two nodes $u\neq v$ with indegree zero. Then, since there is no path from $u$ to $v$, the length of the maximum path is at most $(n+1)-1-1 < n$. We now construct two distinct topological orderings. Since $v$ has indegree zero, we choose $v$ to be the first node. As in the previous case, we can build a topological ordering on $D$ by appending a topological ordering on $D_v$, the subgraph of $D$ over $V\setminus \{v\}$. In any such topological ordering, $v$ comes before $u$. Similarly, since $u$ has indegree zero, we can choose $u$ to be the first node, and complete the topological ordering over $D$ by appending a topological ordering over $D_u$, the subgraph of $D$ over $V\setminus\{u\}$. in any such topological ordering, $u$ comes before $v$. Hence there are at least 2 topological orderings of $D$.
\end{lemmaproof}

\begin{lemma}[NP-Hardness of Maximum Independent Matching]
\label{lem:nphmatching}
	Let $G(V,E)$ be given. It is NP-Hard to calculate the size of the maximum independent matching in $G$.
\end{lemma}

\begin{lemmaproof}
	By reduction to maximum independent set. Let $G(V,E)$ be given. We know it is NP-hard to find the size of its maximum independent set. Consider the graph $G'(Z,E')$ over $Z = V\cup W$ where $W$ is distinct set of nodes such that $|W|=|V|$. We label the nodes both in $V$ and $W$ such that $V=\{v_1,\dots,v_n\}$ and $W=\{w_1,\dots,w_n\}$. The set of edges $E'$ is as follows. For each edge $(v_i,v_j)\in E$, we create four edges in $E'$: $(v_i,v_j)$, $(v_i,w_j)$, $(w_i,w_j)$ and $(w_i,v_j)$. Hence $|Z| = 2|V|$ and $|E'| = 4|E|$. 
	
	We now prove that the size of the maximum independent matching in $G'$ is the same as the size of the maximum independent set in $G$. Let ${\cal I}(G)$ and ${\cal I}(G')$ denote the size of the maximum independent set of $G$ and $G'$ respectively. Note that $G$ is a subgraph of $G'$, hence ${\cal I}(G)\leq {\cal I}(G')$. Let $I,J\subseteq \{1,\dots,n\}$ such that $\{v_i,w_j|\ i\in I,\ j\in J\}$ is a maximum independent set of $G'$. Since for all $i$, $(v_i,w_i)\in E'$, it follows that $I\cap J = \emptyset$. Also, $\Gamma_{G'}(v_i)\cup \{v_i\} = \Gamma_{G'}(w_i)\cup \{w_i\}$. Hence $\{v_i|\ i\in I\cup J\}$ is also an independent set of $G'$. This proves ${\cal I}(G)={\cal I}(G')$.
	
	Now let ${\cal MI}(G')$ be the size of the maximum independent matching in $G'$. It is clear that ${\cal MI}(G')\leq {\cal I}(G')$ as selecting only one of the endpoints of the edges in any independent matching creates an independent set. Let again $\{v_i|\ i\in I\}$ be a maximum independent set in $G'$ (that we assumed without loss of generality to be composed of nodes only in $V$). Since for any $J\subseteq \{1,\dots,n\}$, $\{(v_j,w_j)|\ j\in J\}$ is a matching in $G'$, it follows that $\{(v_i,w_i)|\ i\in I\}$ is an independent matching in $G'$ of size $|I| = {\cal I}(G')$. Hence ${\cal MI}(G') = {\cal I}(G')$. But ${\cal I}(G') = {\cal I}(G)$, which yields ${\cal MI}(G') = {\cal I}(G)$.
	
	Since we can construct $G'$ in linear time, finding the size of the maximum independent matching is as hard as finding the size of the maximum independent set.
\end{lemmaproof}

\end{document}